\documentclass[12pt,a4paper]{article}
\usepackage{epsfig}
\begin{document}
\title{Littlest Higgs model and associated ZH production at high
energy $e^{+}e^{-}$ collider }
\author{Chongxing Yue$^{a}$, Shunzhi Wang$^{b}$, Dongqi Yu$^{a}$\\
{\small $^{a}$ Department of Physics, Liaoning Normal University,
Dalian 116029, China}\thanks{E-mail:cxyue@lnnu.edu.cn}
\\ {\small $^{b}$ College of Physics and Information
Engineering,}\\
\small{Henan Normal University, Xinxiang  453002, China} }

\date{\today}
\maketitle
\begin{abstract}
\hspace{5mm} In the context of the littlest Higgs (LH) model, we
consider the Higgs strahlung process $e^{+}e^{-}\rightarrow ZH $.
We find that the correction effects on this process mainly come
from the heavy photon $B'$. If we take the mixing angle parameter
$c$ in the range of 0.75 - 1, the contributions of the heavy gauge
boson $W_{3}'$ is larger than $6\%$. In most of the parameter
space, the deviation of the total production cross section
$\sigma^{tot}$ from its SM value is larger than $5\%$, which may
be detected in the future high energy $e^{+}e^{-}$ collider (LC)
experiments. The future LC experiments could test the LH model by
measuring the cross section of the process $e^{+}e^{-}\rightarrow
ZH $. \vspace{1cm}

PACS number: 12.60.Cn, 14.80.Bn, 14.70.Hp
\end {abstract}

\newpage

\vspace{.5cm} \noindent{\bf 1. Introduction}

 The standard model (SM) accommodates fermion and weak gauge boson
 masses by including a fundamental scalar Higgs H. However, the SM
 can not explain the dynamics responsible for the generation of
 mass. Furthermore, the scalar sector suffers from the problems of
 triviality and unnaturalness. Thus, the SM can only be an
 effective field theory below some high-energy scale. New physics
 should exist at energy scales around $TeV$. The possible new physics
 scenarios at the $TeV$ scale might be supersymmetry \cite{y1}, dynamical
  symmetry breaking \cite{y2}, extra dimensions \cite{y3}. The present and
 future high energy collider experiments will test these scenarios
 and tell us which might be correct.

 Recently, significant attention has been paid to the class of models of
 electroweak symmetry breaking , known as ``little Higgs models''[4,5,6].
 They provide a way to stabilize the weak scale from the radiative
 corrections of the SM and an alternative to traditional
 candidates for new physics at the $TeV$ scale.They explain how the SM could
 be embedded in a theory valid beyond $1 TeV$, which might solve the
 problems arising from the scalar Higgs boson in the SM. Little
 Higgs models employ an extended set of global and gauge symmetries in
 order to avoid the one-loop quadratic divergences. In  little
 Higgs models, the Higgs boson is a pseudo-Goldstone boson which is
 kept light by an approximate global symmetry and free from
 one-loop quadratic sensitivity to the cutoff scale $\Lambda_{S}$. In
 general, these kinds of models predict the existence of the new
 heavy gauge bosons, such as $W'^{\pm}, W_{3}'$ and $B'$ in the
 extended gauge sector, which can cancel the quadratic divergences
 from the gauge interactions in the SM. These new particles may
 have significant contributions to the low energy observables and
 thus  the precision measured data can give severe constraints on the
 free parameters of these kinds of models \cite{y7,y8,y9}.

  As the simplest realization of the little
 Higgs idea, the littlest Higgs (LH) model \cite{y5} is the
 smallest extension of the SM to date which stabilizes the
 electroweak scale and remains weakly coupled at $TeV$ scale. The LH
 model consists of an $SU(5)$ non-linear $\sigma$ model which is
 broken down to $SO(5)$ via a vacuum expectation value (VEV) of order
 $f$. The subgroup $[SU(2)\times U(1)]^{2}$ of $SU(5)$ is promoted to a
 local gauge symmetry which is broken at the same time to its
 diagonal subgroup $SU(2)\times U(1)$, identified as the SM
 electroweak gauge group. The LH model predicts the existence of
 the new heavy particles, such as $W'^{\pm}$, $W_{3}'$ and $B'$,
 which should not be much heavier than $1TeV$. The
 characteristic signatures of the LH model at the
 present and future collider experiments and the production and
 decay of these new particles have been  studied in Refs.\cite{y8,y10,y11}.
 In this paper, we consider the contributions of these new particles
 to associated ZH production at high energy linear
 $e^{+}e^{-}$ collider (LC) experiments.

 The next generation of LC is expected to operate at energies from
 $300 GeV$ up to about $1 TeV$ \cite{y12}. The Higgs strahlung process
 $e^{+}e^{-} \rightarrow ZH$ is one of the dominant production
 mechanism of the Higgs boson in the future LC experiments. For
 the centre-of-mass energy $\sqrt{\tilde{s}}=350 GeV $ and $500 GeV$ and
 an integrated luminosity of $500 fb^{-1}$, this process ensures
 the observation of Higgs up to the production kinematical limit
 independently of its decay \cite{y13}. In this paper, we
 calculate the cross section of the process $e^{+}e^{-} \rightarrow
 ZH $ in the LH model. Comparing the process $e^{+}e^{-} \rightarrow
 ZH $ in the SM, this process in the LH model receives the
 additional contributions arising from the new gauge bosons $W_{3}'$ and
 $B'$. We find that the new particles $W_{3}'$ and $B'$ can
 significant vary the production cross section of the process
 $e^{+}e^{-} \rightarrow ZH$. In most of the parameter space of the
 LH model, the deviation of the total production cross section
 from its SM value is larger than $5 \%$. The future LC experiments
 may detect the correction effects and further test the LH model.

 In the rest of this paper, we give our results in detail. The
 couplings of the new gauge bosons $B'$ and $W_{3}'$ to ordinary
 particles are given in Sec.2, which are related to our calculation. The
 contributions of these new particles to associated ZH
 production are calculated in Sec.3. Our conclusions are given in
 Sec.4.

\vspace{0.5cm}
 \noindent{\bf 2. The relative couplings of the neutral gauge bosons to
  ordinary particles }

The LH model \cite{y5} is embedded into a non-linear $\sigma$
model with the coset space of $ SU(5)/SO(5)$. At the scale
$\Lambda_{S} \sim 4\pi f $, the global $ SU(5)$ symmetry is broken
into its subgroup $ SO(5)$ via a VEV of order $f$, resulting in
$14 $ Goldstone bosons. The effective field theory of these
Goldstone bosons is parameterized by a non-linear $\sigma$ model
with gauge symmetry $[SU(2)\times U(1)]^{2}$, spontaneously broken
down to the SM gauge group $SU(2)\times U(1)$. The gauge fields
$W^{'\mu}$ and $B^{'\mu}$ associated with the broken gauge
symmetries are related with the SM gauge fields by:
\begin{equation}
W=sW_{1}+cW_{2},\hspace{2cm} W^{'}=-cW_{1}+sW_{2},
\end{equation}
\begin{equation}
B=s^{'}B_{1}+c^{'}B_{2},\hspace{2cm} B^{'}=-c^{'}B_{1}+s^{'}B_{2},
\end{equation}
 with the mixing angles of
$$c=\frac{g_{1}}{\sqrt{g_{1}^{2}+g_{2}^{2}}},\hspace{2.5cm}
c^{'}=\frac{g_{1}^{'}}{\sqrt{g_{1}^{'2}+g_{2}^{'2}}}.$$
 The SM gauge couplings are $g=g_{1}s=g_{2}c$ and
$g^{'}=g_{1}^{'}s^{'}=g_{2}^{'}c^{'}$. In our calculation, we will
take the mass scale $f$, the mixing angles $c$ and $c^{'}$ as free
parameters.

  We denote the SM gauge boson mass eigenstates as $W^{\pm}$, $Z $ and $A $
  and the new heavy gauge boson mass eigenstates as $W'^{\pm}$, $W_{3}'$ and
$B'$. The neutral gauge boson masses are given to leading order
by[8]:
\begin{eqnarray}
M_{A }^{2}=0,\hspace{0.5cm}
M_{B'}^{2}=(M_{Z}^{SM})^{2}S_{W}^{2}(\frac{f^{2}}{5s^{'2}c^{'2}\nu^{2}}-1
+\frac{\chi_{H}C_{W}^{2}}{4s^{2}c^{2}S_{W}^{2}}),
\end{eqnarray}
\begin{eqnarray}
M_{Z }^{2}=(M_{Z}^{SM})^{2}\{1-\frac{\nu^{2}}{f^{2}}[\frac{1}{6}
+\frac{1}{4}(c^{2}-s^{2})^{2}+
\frac{5}{4}(c^{'2}-s^{'2})^{2}+\frac{\chi^{2}}{2}]\},
\end{eqnarray}
\begin{eqnarray}
M_{W_{3}'}^{2}=(M_{Z}^{SM})^{2}C_{W}^{2}(\frac{f^{2}}{s^{2}c^{2}\nu^{2}}-1
-\frac{\chi_{H}S_{W}^{2}}{s^{'2}c^{'2}C_{W}^{2}}),
\end{eqnarray}
with
$$\chi=\frac{4f\nu'}{\nu^{2}}, \hspace{1cm}
\chi_{H}=\frac{5S_{W}C_{W}}{2}\frac{scs^{'}c^{'}(c^{2}s^{'2}
+s^{2}c^{'2})}{5C_{W}^{2}s^{'2}c^{'2}-S_{W}^{2}s^{2}c^{2}}.$$
Where $ \nu = 246 GeV $ is the electroweak scale, $ \nu'$ is the
vacuum expectation value of the scalar $SU(2)_{L}$ triplet and $
\theta_{W} $ is the Weinberg angle. The parameter $ \chi < 1 $
parameterizes the ratio of the triplet and doublet VEV's. In the
following calculation, we will take $ \chi =0.5 $. From above
equations, we can see that the mass $M_{Z}^{SM}$ of the SM gauge
boson Z gets a correction at order $\frac{\nu^{2}}{f^{2}}$. Since
the final $U(1)_{QED}$ symmetry remains intact, the mass and
couplings of the photon are the same as those in the SM. For $f <
3 TeV$, the mass of the heavy photon $B'$ may be lighter than $500
GeV$ \cite{y11}. In most of the parameter space of the LH model,
the mass of the heavy gauge boson $W_{3}'$ is in the range of
$1\sim 3 TeV$.

The couplings of the neutral gauge bosons to the Higgs boson and
charged leptons can be written as:
\begin{equation}
g_{L}^{Z l\bar{l}}=\frac{e}{S_{W}C_{W}}\{(-\frac{1}{2}+S_{W}^{2})
         +\frac{\nu^{2}}{f^{2}}[\frac{c^{2}}{2}(c^{2}-\frac{1}{2})
         -\frac{5}{4}(2c^{'2}-1)(c^{'2}-\frac{2}{5})]\},
\end{equation}

 \begin{equation}
g_{R}^{Z l\bar{l}}=\frac{e}{S_{W}C_{W}}[S_{W}^{2}
         +\frac{5}{2}\frac{\nu^{2}}{f^{2}}(2c^{'2}-1)(c^{'2}-\frac{2}{5})],
\end{equation}

 \begin{equation}
g_{L}^{W_{3}'l\bar{l}}=\frac{e}{2S_{W}}\frac{c}{s},\hspace{1cm}
g_{R}^{W_{3}'l\bar{l}}=0,
\end{equation}

\begin{equation}
g_{L}^{B'l\bar{l}}=\frac{e}{2C_{W}s^{'}c^{'}}(c^{'2}-\frac{2}{5}),\hspace{1cm}
g_{R}^{B'l\bar{l}}=\frac{e}{C_{W}s^{'}c^{'}}(c^{'2}-\frac{2}{5}),
\end{equation}

\begin{equation}
g^{HZ_{\mu}Z_{\nu}}=\frac{ie^{2}\nu
g_{\mu\nu}}{2S_{W}^{2}C_{W}^{2}}\{1-\frac{\nu^{2}}{f^{2}}
[\frac{1}{3}-\frac{3}{4}\chi^{2}+\frac{1}{2}(c^{2}-s^{2})^{2}
+\frac{5}{2}(c^{'2}-s^{'2})^{2}]\},
\end{equation}

\begin{equation}
g^{HZ_{\mu}W_{3\nu}'}=-\frac{ie^{2}\nu
g_{\mu\nu}}{2S_{W}^{2}C_{W}}\frac{(c^{2}-s^{2})}{2sc},
\end{equation}

\begin{equation}
g^{HZ_{\mu}B'_{\nu}}=-\frac{ie^{2}\nu
g_{\mu\nu}}{2S_{W}C_{W}^{2}}\frac{(c^{'2}-s^{'2})}{2s^{'}c^{'}}.
\end{equation}
Where $l$ respects the charged lepton $e, \mu$ or $\tau$. If we
ignore the final state masses, the partial decay widths of the
heavy SU(2) gauge bosons $V'(V=W_{3},W^{\pm})$ can be written as
\cite{y8,y10}:
\begin{equation}
\Gamma(V'\rightarrow
f^{'}\bar{f^{'}})=\frac{C}{24\pi}((g_{L}^{V'f^{'}\bar{f^{'}}})^{2}
+(g_{R}^{V'f^{'}\bar{f^{'}}})^{2})M_{V'},
\end{equation}
\begin{equation}
\Gamma(V'\rightarrow V
H)=\frac{g^{2}\cot^{2}2\theta}{192\pi}M_{V'}=\frac{\alpha
\cot^{2}2\theta}{48S_{W}^{2}}M_{V'},
\end{equation}
where $f^{'}$ is any of the SM quarks or leptons, $C$ is the
fermion color factor and C=1(3) for leptons (quarks). $\theta$ is
the mixing angle between $V'$ and $V $. For the heavy gauge boson
$W_{3}'$, the total decay width is:

\begin{equation}
\Gamma(W_{3}'\rightarrow
total)=\frac{\alpha}{192S_{W}^{2}}[\frac{192c^{2}}{s^{2}}+
\frac{(c^{2}-s^{2})^{2}}{s^{2}c^{2}}]M_{Z'},
\end{equation}
where $\alpha$ is the fine structure constant. Considering the
precision  data constraints, the mass $M_{B'}$ of the heavy photon
$B'$ is not too heavy and is allowed to be in the region of a few
hundred $GeV$[9]. For the decay channels $B'\rightarrow t \bar{t}$
and $B'\rightarrow Z H$, we can not neglect the final state
masses. The possible decay channels of the heavy photon $B'$ have
been discussed in Ref.\cite{y11}.

\vspace{0.5cm}
 \noindent{\bf 3. The process $e^{+}e^{-} \rightarrow ZH$ in the LH model}

 The Higgs strahlung process $e^{+}e^{-} \rightarrow ZH$ is one of
 the dominant production mechanism of the Higgs boson in the LC
 experiments. In the SM , the total cross section of this process
 at leading order is\cite{y14}:
 \begin{equation}
 \sigma^{SM}=\frac{(M_{Z}^{SM})^{4}G_{F}^{2}[1-4S_{W}^{2}+8S_{W}^{4}]}{48\pi}
 \frac{\sqrt{\lambda}(\lambda
 +12\tilde{s}M_{Z}^{2})}{D\tilde{s}^{2}},
 \end{equation}
 where $\sqrt{\tilde{s}} $ is the centre-of-mass energy,
  $\lambda=[\tilde{s}-(M_{Z}+M_{H})^{2}][\tilde{s}-(M_{Z}-M_{H})^{2}]$ and
  $D=(\tilde{s}-M_{Z}^{2})^{2}+M_{Z}^{2}\Gamma_{Z}^{2}$.

Compared the process $e^{+}e^{-} \rightarrow ZH$ in the SM, this
process receives additional contributions from the heavy gauge
bosons $W_{3}'$ and $B'$ in the LH model. Furthermore, in the LH
model, the couplings of the SM gauge boson Z to electrons are
corrected at the order of $\frac{\nu^{2}}{f^{2}}$. The
interference effects between the correction terms and the
tree-level SM coupling terms can also produce corrections to the
production cross section of the process $e^{+}e^{-} \rightarrow
ZH$ at the order of $\frac{\nu^{2}}{f^{2}}$, which are of the same
order as the corrections induced by $W_{3}'$ exchange. Using
Eq.(6)
---Eq.(12), we can give the total production cross section
$\sigma^{tot}$ of this process in the LH model:
\begin{eqnarray}
\sigma^{tot}&=&
               \frac{M_{Z }^{4}G_{F}^{2}}{48\pi s^{4}s^{'4}c^{'4}}
               \{s^{4}s^{'4}c^{'4}(1-2a)[(8C_{W}^{4}-12C_{W}^{2}+5)\\\nonumber
               &-&4(\nu^{2}/f^{2})(C_{W}^{2}-0.5)c^{2}(c^{2}-0.5)
               -20(\nu^{2}/f^{2})(C_{W}^{2}-1.5)(c^{'4}-0.9c^{'2}+0.2)]/D_{Z }\\\nonumber
            &+&C_{W}^{4}s^{'4}c^{'4}(c^{2}-0.5)^{2}/D_{W_{3}'}
            +5S_{W}^{4}s^{4}(c^{'4}-0.9c^{'2}+0.2)/D_{B'}
            \\ \nonumber
            &+&2C_{W}^{2}s^{2}s^{'4}c^{'4}(1-a)(c^{2}-0.5)\\ \nonumber
            &\cdot&[(C_{W}^{2}-0.5)-(\nu^{2}/2f^{2})c^{2}(c^{2}-0.5)
            +(5\nu^{2}/2f^{2})(c^{'4}-0.9c^{'2}+0.2)]/D_{Z W_{3}'}\\ \nonumber
            &+&2S_{W}^{2}s^{4}s^{'2}c^{'2}(1-a)(c^{'4}-0.9c^{'2}+0.2)\\ \nonumber
            &\cdot&[(3C_{W}^{2}-2.5)-(\nu^{2}/2f^{2})c^{2}(c^{2}-0.5)
            -(15\nu^{2}/2f^{2})(c^{'4}-0.9c^{'2}+0.2)]/D_{Z B'}\\ \nonumber
            &+&S_{W}^{2}C_{W}^{2}s^{2}s^{'2}c^{'2}(c^{2}-0.5)
            (c^{'4}-0.9c^{'2}+0.2)/D_{W_{3}'B'}\}\\ \nonumber
            &.&\frac{\sqrt{\lambda}(\lambda+12\tilde{s}(M_{Z}^{SM})^{2})}
            {\tilde{s}^{2}}.
\end{eqnarray}
Where
$$a=\frac{\nu^{2}}{f^{2}}
[\frac{1}{3}-\frac{3}{4}\chi^{2}+\frac{1}{2}(c^{2}-s^{2})^{2}
+\frac{5}{2}(c^{'2}-s^{'2})^{2}]$$
$$D_{V_{i}}=(\tilde{s}-M_{V_{i}}^{2})^{2}+M_{V_{i}}^{2}\Gamma_{V_{i}}^{2},$$
$$D_{V_{i}V_{j}}=\frac{[(\tilde{s}-M_{V_{i}}^{2})^{2}+M_{V_{i}}^{2}
\Gamma_{V_{i}}^{2})]
                 [(\tilde{s}-M_{V_{j}}^{2})^{2}+M_{V_{j}}^{2}
                 \Gamma_{V_{j}}^{2})]}
                 {2[(\tilde{s}-M_{V_{i}}^{2})(\tilde{s}
                 -M_{V_{j}}^{2})+M_{V_{i}}M_{V_{j}}\Gamma_{V_{i}}
                 \Gamma_{V_{j}}]}.$$
In above equations, $V_{i}$ is $Z ,W_{3}'$ or $B'$ and
$\Gamma_{V_{i}}$ is the total width of the gauge boson $V_{i}$.

To obtain numerical results, we take $\alpha = \frac{1}{128.8},
S_{W}^{2}=0.2315, M_{Z}^{SM}=91.18 GeV$ and $\Gamma_{Z}=2.49 GeV$
\cite{y15}. Normalized to the SM cross section $\sigma^{SM}$, the
production cross section of the process $e^{+}e^{-} \rightarrow
ZH$ in the LH model is almost independent of the Higgs boson mass
$M_{H}$ because of the near cancellation of the $M_{H}-$
dependence of the production cross section between in the SM and
in the LH model. Thus, in our numerical calculation, we will
assume $\sqrt{\tilde{s}}=500 GeV, M_{H}=120 GeV$ and take $c,
c^{'}$ and $f$ as free parameters.

The relative correction $\frac{\sigma^{tot}}{\sigma^{SM}}$ is
plotted in Fig.1 as a function of the mixing angle parameter
$c^{'}$ for $f=2 TeV$ and three values of the mixing angle
parameter $c$. From Fig.1 we can see that the relative correction
$\frac{\sigma^{tot}}{\sigma^{SM}}$ is not sensitive to the mixing
angle parameter $c$ for $c< 0.8$. This means that the
contributions of the new particles to the process $e^{+}e^{-}
\rightarrow ZH$ mainly come from the heavy photon $B'$ in most of
the parameter space. This is because the heavy gauge boson
$W_{3}'$ mass square $M_{W_{3}'}^{2}$ is larger than that of the
heavy photon $B'$ at least by an order of magnitude[11]. When $c=
\frac{1}{\sqrt{2}}$, the $W_{3}'$ has no contributions to this
process because the couplings of $W_{3}'$ to the gauge boson $B'$
and the SM Higgs H vanish. In this case, the deviation of the
total cross section $\sigma^{tot}$ from its SM value is larger
than $5 \%$ in most of the parameter space. However, for $0.8 \leq
c <1$, the contributions of the $W_{3}'$ can not be ignored. The
absolute value of the
$\frac{\sigma^{tot}-\sigma^{SM}}{\sigma^{SM}}$ is larger than $10
\% $, which might be detected in the future LC experiments.

\vspace*{6.4cm}
\begin{figure}[ht]
\begin{center}
\begin{picture}(200,100)(0,0)
\put(-100,10){\epsfxsize130mm\epsfbox{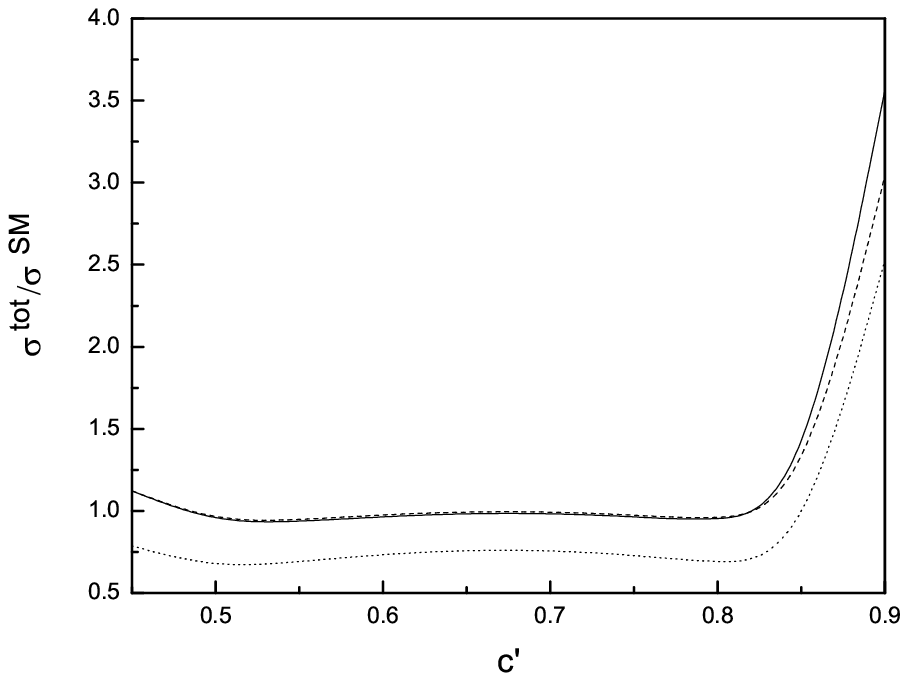}}
\put(-95,10){Fig.1\hspace{5mm}The relative correction
$\sigma^{tot}/\sigma^{SM}$ as a function of $c^{'}$ for $f
 = 2TeV$} \put(-95,-10){
 and c=0.1(solid
line), $\frac{1}{\sqrt{2}}$(dashed line) and 0.9(dotted line).}
\end{picture}
\end{center}
\end{figure}

\vspace*{5.5 cm}
\begin{figure}[ht]
\begin{center}
\begin{picture}(200,60)(0,0)
\put(-100,-40){\epsfxsize130mm\epsfbox{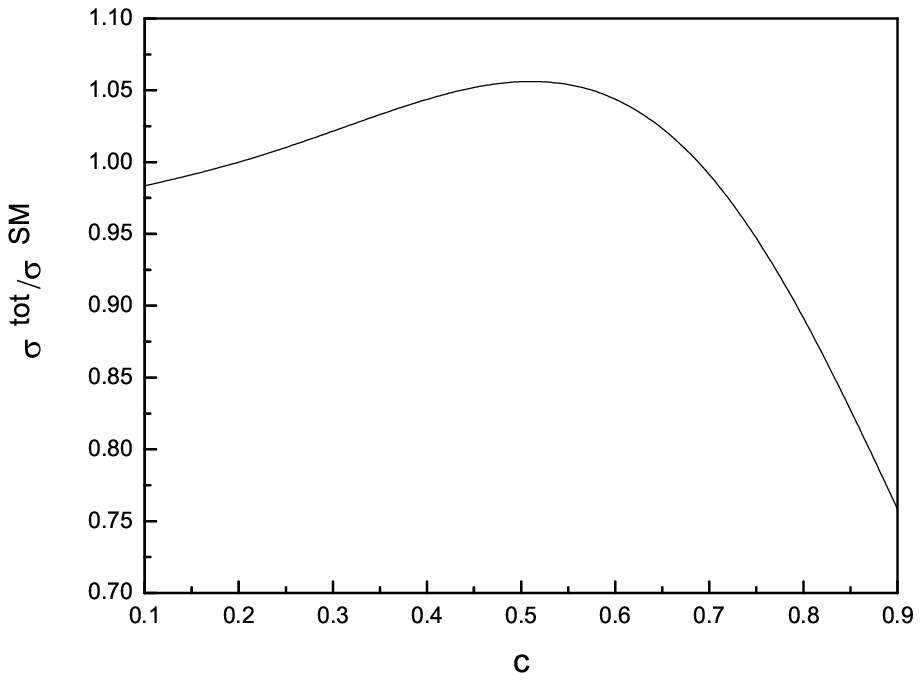}}
\put(-95,-30){Fig.2\hspace{5mm}The relative correction
$\sigma^{tot}/\sigma^{SM}$ as a function of c for $ f = 2TeV$}
\put(-85,-50){
 and $c^{'}=\frac{1}{\sqrt{2}}$.}
\end{picture}
\end{center}
\end{figure}
\vspace*{0.9cm}

The strongest constraints on the mass and couplings of the heavy
photon $B'$ arise from the lack of observation for the production
of $B'$. For example, Ref.[9] has shown that for the global
symmetry parameter $f=2TeV$, there must be $c'<0.24$, which comes
from direct searches at the Tevatron. However, in the modified
version of the LH model\cite{y7}, only one U(1) is gauged, and
there would be no heavy photon $B'$ which corresponds to
$c'=\frac{1}{\sqrt{2}}$. In this case, the LH model avoids
constraints from Tevatron searches for heavy gauge bosons and the
limits on the scale $f$ from the electroweak data are relaxed.

To see the effects of the heavy gauge boson $W_{3}'$ on the
process $e^{+}e^{-} \rightarrow ZH$, we plot the relative
correction $\sigma^{tot}/\sigma^{SM}$ as a function of the mixing
angle parameter $c$ for $f=2 TeV, c^{'}=\frac{1}{\sqrt{2}}$ in
Fig.2. In this case, the contributions of the heavy photon $B'$
vanish. The absolute value of
$\sigma^{tot}-\sigma^{SM}/\sigma^{SM}$ is smaller than $5 \% $ for
$c < 0.7 $. If we assume the mixing angle parameter $ c > 0.75$,
then the gauge boson $W_{3}'$ decreases the cross section of this
process in the SM. The varying value of the cross section,
compared to that in the SM, is larger than $6 \%$.

\vspace*{8.2cm}
\begin{figure}[ht]
\begin{center}
\begin{picture}(200,-100)(0,0)
\put(-100,-40){\epsfxsize130mm\epsfbox{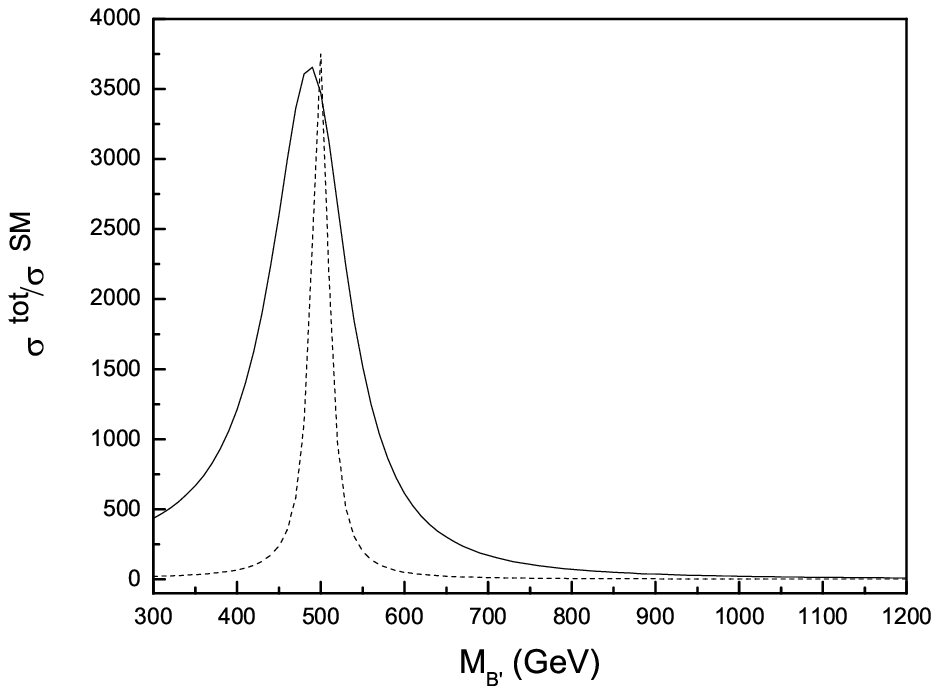}}
\put(-95,-30){Fig.3\hspace{9mm}The relative correction
$\sigma^{tot}/\sigma^{SM}$ as a function of the heavy photon}
\put(-85,-50){mass $M_{B'}$ for $c=\frac{1}{\sqrt{2}}$ and
$c^{'}$=0.1(solid line), $0.2$(dashed line). }
\end{picture}
\end{center}
\end{figure}
 \vspace*{2.1cm}

From Eq.(3) we can see that mass $M_{B'}$ of the heavy photon $B'$
mainly depends on the global symmetry breaking scale $f$ and the
mixing angle parameter $c'$ between two U(1) gauge bosons, while
is insensitive to the value of the mixing angle parameter $c$
between two SU(2) gauge bosons. To further explain the
contributions of $B'$ to associated $ZH$ production, we plot the
relative correction $\sigma^{tot}/\sigma^{SM}$ as a function of
$M_{B'}$ for $c=\frac{1}{\sqrt{2}}$ and two values of the mixing
angle parameter $c^{'}=0.1$(solid line), $0.2$(dashed line). In
this case the contributions of the new particle $W_{3}'$ predicted
by the LH model to the process $e^{+}e^{-} \rightarrow ZH$ is zero
. For the mixing angle parameter $c^{'}=0.1( 0.2 )$, the peak of
the total cross section $\sigma^{tot} $ emerges when the heavy
photon mass approximately equals $ 480GeV( 500GeV)$. Even we take
the heavy photon mass $M_{B'}=1200GeV $, we have
$\sigma^{tot}/\sigma^{SM} = 2 $. Thus, in a sizable parameter
region of the LH model, the heavy photon $B'$ can produce
significant new signal, which can be detected in the future LC
experiments.

The cross section of the process $e^{+}e^{-} \rightarrow ZH$ can
be measured by analysing the mass spectrum of the system recoiling
against the Z boson. For $M_{H}=130 GeV$, the final states are
four jet $b\bar{b}q\bar{q}$ and two jet plus two lepton
$b\bar{b}l^{+}l^{-}$, which are coming from the  Higgs boson
decaying to $b\bar{b}$, the Z boson decaying to a $q\bar{q}$, and
the Z boson decaying to charged leptons, respectively. From the
number of signal events fitted to the di-lepton recoil mass
spectrum, the production cross section of the process $e^{+}e^{-}
\rightarrow ZH$ is obtained with a statistical accuracy $\pm
2.8\%$, combing the $e^{+}e^{-} $ and $\mu^{+}\mu^{-}$ channel
\cite{y12}. In most of the parameter space of the LH model, the
deviation of the total production cross section
 from its SM value is larger than $5 \%$. Even for
 $c^{'}=0.2$, $c=\frac{1}{\sqrt{2}}$, and $M_A'=1200GeV$,
 the value of the $\sigma^{tot}/\sigma^{SM}$ can reach 2.
 Thus, the effects of the new particles predicted by the LH model
might be observable in the future LC experiments.

\vspace{0.5cm} \noindent{\bf 4. Conclusions}

Little Higgs models provide a natural mechanism to cancel
quadratic divergences that appear in the calculation of the Higgs
mass without resorting to supersymmetry. The cancellation of
divergences occurs by the alignment of vacua and the existence of
several new particles. These kinds of models predict the existence
of several scalars, new gauge bosons, and vector-like top quarks.
The possible signatures of these models might be detected in the
future high energy experiments.

The Higgs-strahlung process $e^{+}e^{-} \rightarrow ZH$ is one of
the main production process of the Higgs boson H at the LC
experiments, which offers a very distinctive signature ensuring
the observation of the SM Higgs boson up to the production
kinematical limit independently of its decay. Using this process,
we can precisely measure the Higgs mass $M_{H}$ and the couplings
of Higgs boson to massive gauge bosons and determine the quantum
numbers of the Higgs boson. Thus, it is necessary to consider the
process $e^{+}e^{-} \rightarrow ZH$ in the context of the little
Higgs models and see whether this process can be used to test
these models.

In this paper, we calculate the contributions of the new gauge
bosons $W_{3}', B'$ predicted by the LH model to the cross section
of this process and find that the cross section can be
significantly varied. In most of the parameter space, the
corrections mainly come from the heavy photon $B'$. With
reasonable values of the parameters in the LH model, the deviation
of the total production cross section $\sigma^{tot}$ from its SM
value is larger than $5 \%$. If we assume that the mixing angle
parameter $c$ is in the range of $0.75 - 1 $, the contributions of
the heavy gauge boson $W_{3}'$ to the process $e^{+}e^{-}
\rightarrow ZH$ is larger than $6 \%$. It has been shown that the
modified version of the LH model\cite{y7}, in which only one U(1)
is gauged, can avoid constraints from Tevatron searches for heavy
gauge bosons and the limits on the free parameters from the
electroweak data are relaxed. Thus, it is possible that this
process can be used to detect the signatures of these models.

\vspace{.5cm} \noindent{\bf Acknowledgments}

  We are very grateful to Prof. X. Zhang for bringing little Higgs
models to our attention. C.X.Yue would like to thank H. E. Logan
et al. for pointing out that the effects of the correction terms
about the mass and couplings of the SM gauge boson Z on the
process $e^{+}e^{-} \rightarrow ZH$ can not be neglected. This
work was supported by the National Natural Science Foundation of
China (90203005).
 \vspace{.5cm}

%\newpage

\end{document}